%% file: main.tex
\documentclass[journal,letterpaper,twocolumn]{IEEEtran}
\usepackage{amsfonts}
\usepackage{amssymb}
\usepackage{cite}
\usepackage[cmex10]{amsmath}
\usepackage{float}
\usepackage{color}
\usepackage{stfloats,fancyhdr}

\usepackage{algorithm}
\usepackage{algorithmic}
\usepackage{multirow}
\usepackage{changepage}
\usepackage[normalem]{ulem}

\newtheorem{remark}{Remark}

\IEEEoverridecommandlockouts

\ifCLASSINFOpdf
  \usepackage[pdftex]{graphicx}
  \DeclareGraphicsExtensions{.pdf,.jpeg,.png}
\else
  \usepackage[dvips]{graphicx}
  \DeclareGraphicsExtensions{.eps}
\fi

\usepackage{subfigure}

\usepackage{fancybox,dashbox}

\begin{document}

\title{Short-Packet Two-Way Amplify-and-Forward Relaying
\thanks{Y. Gu, H. Chen, Y. Li and B. Vucetic are with School of Electrical and Information Engineering, The University of Sydney, Sydney, NSW 2006, Australia (email: yifan.gu@sydney.edu.au, he.chen@sydney.edu.au, yonghui.li@sydney.edu.au, branka.vucetic@sydney.edu.au).}
\thanks{L. Song is with School of Electronics Engineering and Computer Science, Peking University, Beijing, China (email: lingyang.song@pku.edu.cn).}
}

\author{Yifan Gu, He Chen, Yonghui Li, Lingyang Song
and Branka Vucetic}



\maketitle

\begin{abstract}
This letter investigates an amplify-and-forward two-way relay network (TWRN) for short-packet communications. We consider a classical three-node TWRN consisting of two sources and one relay. Both two time slots (2TS) scheme and three time slots (3TS) scheme are studied under the finite blocklength regime. We derive approximate closed-form expressions of sum-block error rate (BLER) for both schemes. Simple asymptotic expressions for sum-BLER at high signal-to-noise ratio (SNR) are also derived. Based on the asymptotic expressions, we analytically compare the sum-BLER performance of 2TS and 3TS schemes, and attain an expression of critical blocklength, which can determine the performance superiority of 2TS and 3TS in terms of sum-BLER. Extensive simulations are provided to validate our theoretical analysis. Our results discover that 3TS scheme is more suitable for a system with lower relay transmission power, higher differences between the average SNR of both links and relatively lower requirements on data rate and latency.
\end{abstract}

\begin{IEEEkeywords}
Ultra-reliable and low latency communication, short-packet communication, finite blocklength, two-way relay.
\end{IEEEkeywords}

\IEEEpeerreviewmaketitle

\input{introduction}
\input{sysmodel}

\input{Performance_Analysis}

\input{numerical}

\section{Conclusions}
In this letter, we revisited the performance of a two-way relay network for short-packet communications in the finite blocklength regime. Specifically, we first derived closed-form expressions of average sum-block error rate (BLER) for both two time slots (2TS) and three time slots (3TS) schemes. To gain further insights and analytically compare the performance of 2TS and 3TS, simple asymptotic expressions of sum-BLER at high SNR range were derived for both schemes. We then defined and attained a critical blocklength which can be used to design a TWRN with various delay requirements. We also concluded that 3TS scheme is more suitable for a system with lower information rate, relatively low transmit power at relay, higher differences between the average SNR of both links and less stringent latency requirement.


\ifCLASSOPTIONcaptionsoff
  \newpage
\fi


\bibliographystyle{IEEEtran}
\bibliography{References}

%

\end{document}

%% file: introduction.tex
\section{Introduction}
Unlike the traditional broadband services mainly designed for human-operated mobile terminals, the fifth-generation (5G) cellular systems aim to provide extensive wireless connections between autonomous machines \cite{towards}. Machine-type communications are typically characterized with short burst packets. In some applications, it also requires ultra low latency, such as vehicle-to-vehicle (V2V) communication and tele-robotic surgery \cite{shortpacket,background1,henryref}. To meet the stringent latency requirement, designing short-packet communication with finite blocklength codes is essential for machine-type communications and has attracted considerable research interests recently (see \cite{FBLtheory1,yulin1,yulin3,Qapproximation,ultrareliable} and the references therein).

In wireless communications, the Shannon theory shows the maximum information rate at which an arbitrarily low error rate can be achieved. However, it requires super-long packet length and is not applicable for short-packet communication scenarios. Motivated by this, \cite{FBLtheory1} developed a new fundamental framework for short-packet communications and derived an error probability bound for a given blocklength and coding rate. It is shown that the block error rate (BLER) increases as the blocklength reduces. This new theoretical framework opens a new research direction for the design and analysis of short-packet communications and requires a fundamental revisit of conventional cooperative networks, which are designed based on Shannon theory for extremely large packet length scenarios. Specifically, using a relay can improve the system performance on one hand by providing power gain but it may degrade the system performance on the other hand since the blocklength of each hop is halved compared with a non-cooperative transmission. This trade-off has been studied very recently in \cite{yulin1,yulin3} for one-way relaying network under both static and quasi-static channel conditions. Very recently, \cite{yifan} revisited the fundamental comparison between half-duplex relaying and full-duplex relaying for short-packet communications in the finite blocklength regime.

Compared to one-way relaying, two-way relaying can further enhance the system spectral efficiency and reduce the network latency by allowing two sources to transmit/receive information simultaneously. To the best knowledge of the authors, the performance of two-way relaying under the finite blocklength regime has not been studied in the open literature. Motivated by this gap, in this paper we investigate a two-way relay network (TWRN) wherein two sources $A, B$ exchange information with the help of one amplify-and-forward (AF) relay $R$. The information exchange between $A$ and $B$ can be accomplished using two time slots (2TS) or three time slots (3TS) schemes \cite{Raymond_twowayrelay}. Specifically, in the 2TS scheme, $A$ and $B$ simultaneously transmit information to $R$ during the first time slot. In the second time slot, $R$ amplifies and forwards the received signal to $A,B$. In the 3TS scheme\footnote{The considered 3TS scheme is also known as three-phases analog network coding (ANC) \cite{new}. Another 3TS scheme is time division broadcast (TDBC) \cite{lastref}, which is different from three-phases ANC in terms of scaling gains.}, $A$ and $B$ transmit information to $R$ sequentially during the first and second time slots. In the third time slot, $R$ first processes the received signal and then amplifies and forwards it to $A$ and $B$. It is shown in \cite{Raymond_twowayrelay} that 3TS outperforms 2TS in terms of the sum-bit error rate for various system setups. This is mainly because that in 3TS, the signals from $A$ and $B$ are received separately such that the relay can optimize its power allocation to these two signals to improve the system performance. However, when it comes to short-packet communication scenarios with finite blocklength codes, 3TS scheme has a shorter blocklength for each time slot than 2TS scheme, which could lead to a higher error rate for 3TS. Thus, it is necessary to revisit the performance comparison of 2TS and 3TS schemes of TWRNs in the finite blocklength regime.

It is worth pointing out that in the existing studies of long packet transmissions, the Shannon bound was applied, which only involves a logarithm function and depends on the received SNR. In this case, only the received SNR and the coding rate of the system are needed to be considered in the comparison between 2TS and 3TS schemes \cite{Raymond_twowayrelay}. In contrast, for the considered short-packet communications, we need to jointly consider the received SNR, the coding rate, the blocklength and the BLER to compare these two relaying schemes properly. Furthermore, the new rate bound for short-packet communications has a more complex structure and involves a complicated $Q$-function. As such, the performance analysis and optimization of short-packet communications require more complex mathematical manipulations and new approximation methods.

\textbf{\emph{Notation}}: Throughout this paper, we use $f_{X}(x)$ and $F_{X}(x)$ to denote the probability density function (PDF) and cumulative distribution function (CDF) of a random variable $X$. $\mathbb{E}\left\{ {\cdot} \right\}$ represents the expectation operator. $Q\left( x \right) = \int_x^\infty  {{1 \over {\sqrt {2\pi } }}{e^{ - {{{x^2}} \over 2}}}dx} $ is the complementary Gaussian cumulative distribution function. $K_v \left( {\cdot} \right)$ is the modified Bessel function of the second kind \cite[Eq. (8.432)]{Tableofintegral} and $G_{ \cdot , \cdot }^{ \cdot , \cdot }\left( \cdot \right)$ is the Meijer's G-function \cite[Eq. (9.301)]{Tableofintegral}.

%% file: sysmodel.tex
\section{System Model}
We consider a two-way relay network (TWRN) consisting of two sources, $A$ and $B$, and one amplify-and-forward (AF) relay $R$. All nodes in the considered TWRN are single-antenna and half-duplex devices. We assume that there is no direct link between two sources due to obstacles or severe attenuation, and the information exchange can be only performed via the intermediate relay. Each transmission block has a length of $m$ symbols (i.e., channel uses). The links between $A$, $R$ and $B$, $R$ are assumed to suffer from Rayleigh fading with average power gain $\Omega_A$ and $\Omega_B$, respectively. We assume quasi-static fading channels for which the fading coefficients remain constant during each transmission block and change independently from one block to the other. 

In the following, we denote by $P_A$, $P_B$ and $P_R$ the transmit power of $A$, $B$ and $R$, respectively. We use $N_A$, $N_B$ and $N_R$ to denote the variance of additive white Gaussian noise (AWGN) at $A$, $B$ and $R$, respectively. For notation simplicity, we define the average SNR for the four links as ${\overline\gamma  _{A,R}} = {{{P_A}\Omega_A} \over {{N_R}}}$, ${\overline \gamma  _{B,R}} = {{{P_B}\Omega_B} \over {{N_R}}}$, ${\overline \gamma  _{R,A}} = {{{P_R}\Omega_A} \over {{N_A}}}$ and ${\overline \gamma  _{R,B}} = {{{P_R}\Omega_B} \over {{N_B}}}$. According to \cite{Raymond_twowayrelay}, the exact end-to-end signal-to-noise ratio (SNR) at $A$ and $B$ implementing 2TS or 3TS schemes can be summarized as

\begin{equation}\label{SNR}
{\gamma _{\delta ,\tau}} = {{{c_{\delta ,\tau}}HG} \over {{a_{\delta ,\tau}}G + {b_{\delta ,\tau}}H + 1}},
\end{equation}
where $\delta \in \left\{A,B\right\}$ represents the received SNR at the two sources, $\tau \in \left\{2,3\right\}$ indicates 2TS, 3TS schemes, and $H$, $G$ are exponentially distributed random variables with unit power gain. The parameters ${a_{\delta ,\tau}}$, ${b_{\delta ,\tau}}$ and ${c_{\delta ,\tau}}$ in (\ref{SNR}) are summarized in Table I, in which $\alpha$ and $\beta$ with $\alpha^2+\beta^2=1$ are the power allocation factors for $A$'s signal and $B$'s signal in 3TS \cite{Raymond_twowayrelay}. Note that the optimal values of $\alpha$ and $\beta$ are derived later in the performance analysis section.

\begin{table}[]
\centering
\caption{Parameters}
\label{my-label}
\resizebox{\columnwidth}{!}{%
\begin{tabular}{|l|l|l|l|l|l|l|}
\hline
  & Source $A$, 2TS & Source $B$, 2TS & Source $A$, 3TS & Source $B$, 3TS  \\ \hline
${a_{\delta ,\tau}}$ & ${\overline \gamma  _{R,A}} + {\overline \gamma  _{A,R}} $     & $ {\overline \gamma  _{R,B}}+ {\overline \gamma  _{B,R}}$      & ${\overline\gamma_{R,A}}+\beta^2 {\overline \gamma  _{A,R}} $ &  ${\overline \gamma  _{R,B}} +\alpha^2 {\overline \gamma  _{B,R}}$    \\ \hline

${b_{\delta ,\tau}}$ &  ${\overline \gamma  _{B,R}}  $   &   ${\overline \gamma  _{A,R}} $   &  $\alpha^2{\overline \gamma  _{B,R}} $ & $ \beta^2{\overline \gamma  _{A,R}}$
  \\ \hline
${c_{\delta ,\tau}}$ &  ${\overline \gamma  _{R,A}}{\overline \gamma  _{B,R}}  $   &  ${\overline \gamma  _{R,B}}{\overline \gamma  _{A,R}}$ &
$\alpha^2 {\overline \gamma  _{R,A}}{\overline \gamma  _{B,R}}$   &  $ \beta^2 {\overline \gamma  _{R,B}}{\overline \gamma  _{A,R}}$
 \\ \hline
\end{tabular}}
\end{table}

%% file: Performance_Analysis.tex
\section{Performance Analysis}
In this section, we analyze the performance of the considered TWRN in terms of sum-block error rate (BLER) under the finite blocklength regime. The sum-BLER is defined as the sum of individual BLER at $A$ and $B$. We first derive a closed-form expression of sum-BLER for both 2TS and 3TS schemes\footnote{Note that our analysis of 2TS and 3TS schemes can be readily extended to a 4TS scheme.}. In order to gain further insights, we then derive a simple asymptotic expression for sum-BLER at high SNR range.

\subsection{Sum-BLER}
Let $\varepsilon_{\tau}$ be the sum-BLER of the considered TWRN, where $\tau \in \left\{2,3\right\}$ represents different transmission schemes. For a fair comparison, we consider that both $A$ and $B$ exchange $\sigma $ bits of information over $m$ channel uses in each packet transmission. Thus, the duration of each time slot in 2TS and 3TS is ${m_{\tau}} = m/\tau$, measured by channel uses. The coding rate for each time slot in 2TS and 3TS is given by ${r_{\tau}} = {\sigma  \over {{m_{\tau}}}}$. According to \cite[Eq. (59)]{FBLtheory}, when ${m_{\tau}}$ is sufficiently large (i.e., ${m_{\tau}}>100$), $\varepsilon_{\tau}$ can be tightly approximated as
\begin{equation}\label{sumBEP}
\varepsilon_{\tau} \approx \Psi_{A, \tau}+\Psi_{B, \tau}
\end{equation}
where $\Psi_{\delta, \tau} = \mathbb {E} \left\{Q\left( {{{C\left( {{\gamma _{\delta,\tau}}} \right) - {r_{\tau}}} \over {\sqrt {V\left( {{\gamma _{\delta,\tau}}} \right)/{m_{\tau}}} }}} \right) \right\}$, $\delta \in \left\{A,B\right\}$, $C\left( {{\gamma _{\delta,\tau}}} \right) = {\log _2}\left( {1 + {\gamma _{\delta,\tau}}} \right)$ is the Shannon capacity and $V\left( {{\gamma _{\delta,\tau}}} \right) = \left( {1 - {1 \over {{{\left( {1 + {\gamma _{\delta,\tau}}} \right)}^2}}}} \right){\left( {{{\log }_2}e} \right)^2}$ is the channel dispersion which measures the stochastic variability of the channel relative to a deterministic channel with the same capacity \cite{FBLtheory1}.

In the following, we evaluate the general term $\Psi_{\delta, \tau}$ in order to characterize the sum-BLER in (\ref{sumBEP}). It is intractable to evaluate $\Psi_{\delta, \tau}$ in a closed-form and we are thus motivated to first use a linear approximation of $Q\left( {{{C\left( {{\gamma _{\delta,\tau}}} \right) - {r_{\tau}}} \over {\sqrt {V\left( {{\gamma _{\delta,\tau}}} \right)/{m_{\tau}}} }}} \right) \approx \Xi\left({\gamma _{\delta,\tau}}\right)$ given by \cite{Qapproximation,ultrareliable}
\begin{equation}
\Xi \left( {\gamma _{\delta,\tau}} \right) = \left\{ {
\begin{matrix}
\begin{split}
   &1, \quad \quad \quad \quad \quad\quad\quad\quad\quad\quad\quad\quad {\gamma _{\delta,\tau}} \le  \zeta_{\tau} \\
   &{{1 \over 2}} -{\vartheta_{\tau} \sqrt {{m_{\tau}}}}\left( {{\gamma _{\delta,\tau}} - \theta_{\tau} } \right), \quad   \zeta_{\tau}   < {\gamma _{\delta,\tau}} < \xi_{\tau}   \\
  & 0,  \quad \quad \quad\quad\quad\quad\quad\quad\quad\quad\quad\quad {\gamma _{\delta,\tau}} \ge \xi_{\tau} \\
\end{split}
 \end{matrix} } \right.,
\end{equation}
where $\vartheta_{\tau} = {1 \over {2\pi \sqrt {{2^{2{r_{\tau}}}} - 1} }}$, $\theta_{\tau} = {2^{{r_{\tau}}}} - 1$, $\zeta_{\tau} = \theta_{\tau}  - {1 \over {2\vartheta_{\tau} \sqrt {{m_{\tau}}} }}$ and $\xi_{\tau} =\theta_{\tau}  + {1 \over {2\vartheta_{\tau} \sqrt {{m_{\tau}}} }} $.
With the above approximation, $\Psi_{\delta, \tau}$ can be evaluated as
\begin{equation}\label{app1}
\begin{split}
{\Psi _{\delta ,\tau }} &= \int_0^\infty  {Q\left( {{{C\left( x \right) - {r_\tau }} \over {\sqrt {V\left( x \right)/{m_\tau }} }}} \right)} {f_{{\gamma _{\delta ,\tau }}}}\left( x \right)dx \\
& \approx \int_0^\infty  {\Xi \left( x \right)} {f_{{\gamma _{\delta ,\tau }}}}\left( x \right)dx.
\end{split}
\end{equation}

To further evaluate the above integral, we apply the partial integration theorem and ${\Psi _{\delta ,\tau }}$ can be further evaluated as
\begin{equation}
\begin{split}
{\Psi _{\delta ,\tau }} &\approx \int_0^\infty  {\Xi \left( x \right)} d{F_{{\gamma _{\delta ,\tau }}}}\left( x \right) = \left[ {\Xi \left( x \right){F_{{\gamma _{\delta ,\tau }}}}\left( x \right)} \right]_0^\infty  - \\
&\quad \int_0^\infty  {{F_{{\gamma _{\delta ,\tau }}}}\left( x \right)} d\Xi \left( x \right)\\
& = {\vartheta _\tau }\sqrt {{m_\tau }} \int_{{\zeta _\tau }}^{{\xi _\tau }} {{F_{{\gamma _{\delta ,\tau }}}}\left( x \right)} dx.
\end{split}
\end{equation}

Note that an exact closed-form expression of ${F_{{\gamma _{\delta ,\tau}}}}\left( x \right)$ was given in \cite[Eq. (29)]{Raymond_twowayrelay}. However, due to the complicated structure of ${F_{{\gamma _{\delta ,\tau}}}}\left( x \right)$, we still cannot further solve the above integral. To tackle this, we apply the approximation to the received SNR given in (\ref{SNR}), i.e., ${\gamma _{\delta ,\tau }} = {{{c_{\delta ,\tau }}XY} \over {{a_{\delta ,\tau }}Y + {b_{\delta ,\tau }}X + 1}} \approx {{{c_{\delta ,\tau }}XY} \over {{a_{\delta ,\tau }}Y + {b_{\delta ,\tau }}X}}$, such that the CDF of ${\gamma _{\delta ,\tau }}$ can now be approximated as \cite[Eq. (29)]{Raymond_twowayrelay}. To find a closed-form expression of ${\Psi _{\delta ,\tau}}$, we then use a series representation of $\exp \left( { - {{x\left( {{a_{\delta ,\tau}} + {b_{\delta ,\tau}}} \right)} \over {{c_{\delta ,\tau}}}}} \right)  = \sum\limits_{k = 0}^\infty  {{{{{ - {{x\left( {{a_{\delta ,\tau}} + {b_{\delta ,\tau}}} \right)} \over {{c_{\delta ,\tau}}}}}^k}} \over {k!}}} $ \cite[Eq. (1.211-1)]{Tableofintegral}. The term $\Psi_{\delta, \tau}$ can now be evaluated from (\ref{app1}) as

\begin{equation}
\begin{split}
{\Psi _{\delta ,\tau}} &\approx 1-{{{\vartheta _{\tau}}\sqrt {{m_{\tau }}} {c_{\delta ,\tau}}} \over {2\sqrt {{a_{\delta ,\tau}}{b_{\delta ,\tau}}} }}\sum\limits_{k = 0}^\infty  {{{{{\left( { - {{{a_{\delta ,\tau}} + {b_{\delta ,\tau }}} \over {2\sqrt {{a_{\delta ,\tau}}{b_{\delta ,\tau}}} }}} \right)}^k}} \over {k!}}} \\
& \quad \times \int\limits_{{{2\sqrt {{a_{\delta ,\tau}}{b_{\delta ,\tau}}} } \over {{c_{\delta ,\tau}}}}{\zeta _{\delta ,\tau}}}^{{{2\sqrt {{a_{\delta ,\tau}}{b_{\delta ,\tau}}} } \over {{c_{\delta ,\tau}}}}{\xi _{\delta ,\tau}}} {{y^{k + 1}}} {K_1}\left( y \right)dy.\\
\end{split}
\end{equation}
To solve the above integral, we represent the function $y^{k+1} K_1\left(y\right)$ in terms of Meijer's G-function \cite[Eq. (9.34-3)]{Tableofintegral}. After some tedious mathematical manipulations, the above integral involving Meijer's G function can be solved by using the results in \cite{MG} and ${\Psi_{\delta ,\tau}}$ can be expressed as

\begin{equation}\label{generalterm}
\begin{split}
 \Psi_{\delta, \tau} & \approx 1 - {{{\vartheta _{\tau }}\sqrt {{m_{\tau}}} {c_{\delta ,\tau}}} \over {2\sqrt {{a_{\delta ,\tau}}{b_{\delta ,\tau}}} }}\sum\limits_{k = 0}^{{\infty}} {{{{{\left( { - {{{a_{\delta ,\tau}} + {b_{\delta ,\tau}}} \over {\sqrt {{a_{\delta ,\tau}}{b_{\delta ,\tau}}} }}} \right)}^k}} \over {k!}}} \times \\
  &\quad \left[ {\Phi \left( {{{{a_{\delta ,\tau}}{b_{\delta ,\tau}}} \over {{c_{\delta ,\tau}}^2}}{\xi _{\delta ,\tau}}^2} \right) - \Phi \left( {{{{a_{\delta ,\tau TS}}{b_{\delta ,\tau}}} \over {{c_{\delta ,\tau}}^2}}{\zeta _{\delta ,\tau}}^2} \right)} \right],\\
\end{split}
\end{equation}

where
\begin{equation}
\Phi \left( x \right) = G_{1,3}^{2,1}\left( {x\left| {
\begin{matrix}
   1  \\
   {{{k + 3} \over 2},{{k + 1} \over 2},0}  \\

 \end{matrix}
 } \right.} \right).
\end{equation}

Although the expression given in (\ref{generalterm}) has infinite summation terms, we will later show in the simulation that very accurate results can be obtained even by choosing the first two summation terms. The sum-BLER of the considered TWRN can be obtained by substituting (\ref{generalterm}) into (\ref{sumBEP}).

It is worth noting that no closed-form expressions for system performance were characterized in \cite{yulin3} for the one-way relay system, and our analytical expressions can be easily simplified and extended to cover this gap. Due to the complex structure of (\ref{sumBEP}), we cannot further optimize the power allocation factors $\alpha, \beta$ and gain insights on the impacts of various system parameters. We are thus motivated to seek for a simple expression at high SNR range.

\subsection{High SNR}
By using the asymptotic result given in \cite[Eq. (33)]{Raymond_twowayrelay} and the expression given in (\ref{app1}), the asymptotic sum-BLER for the considered 2TS and 3TS schemes at high SNR range can be expressed as
\begin{equation}\label{sumBEPhighSNR}
\varepsilon _{\tau}^\infty  = \left( {{{{a_{A,\tau}} + {b_{A,\tau}}} \over {{c_{A,\tau}}}} + {{{a_{B,\tau}} + {b_{B,\tau}}} \over {{c_{B,\tau}}}}} \right)\left({2^{{{\tau \sigma } \over m}}}-1\right).
\end{equation}

With the above asymptotic expression and the parameters defined in Table I, we can now characterize the optimal values of ${\alpha ^2}$ and ${\beta ^2}$. This can be done by first substitute ${\beta ^2} = 1 - {\alpha ^2}$ into (\ref{sumBEPhighSNR}) and then taking the derivative of (\ref{sumBEPhighSNR}) with respect to ${\alpha ^2}$ and solving the equation ${{d\varepsilon _\tau ^\infty } \over {d{\alpha ^2}}} = 0$. We then obtain that the optimal value of power allocation factors given by ${\alpha ^2} = {{\sqrt {{{{{\bar \gamma }_{A,R}}^2} \over {{{\bar \gamma }_{R,A}}}}{\rm{ + }}{{\bar \gamma }_{A,R}}} } \over { {\sqrt {{{{{\bar \gamma }_{A,R}}^2} \over {{{\bar \gamma }_{R,A}}}}{\rm{ + }}{{\bar \gamma }_{A,R}}} {\rm{ + }}\sqrt {{{{{\bar \gamma }_{B,R}}^2} \over {{{\bar \gamma }_{R,B}}}}{\rm{ + }}{{\bar \gamma }_{B,R}}} }}}$ and ${\beta ^2} = {{\sqrt {{{{{\bar \gamma }_{B,R}}^2} \over {{{\bar \gamma }_{R,B}}}}{\rm{ + }}{{\bar \gamma }_{B,R}}} }\over { {\sqrt {{{{{\bar \gamma }_{A,R}}^2} \over {{{\bar \gamma }_{R,A}}}}{\rm{ + }}{{\bar \gamma }_{A,R}}} {\rm{ + }}\sqrt {{{{{\bar \gamma }_{B,R}}^2} \over {{{\bar \gamma }_{R,B}}}}{\rm{ + }}{{\bar \gamma }_{B,R}}} } }}$.

Moreover, at high SNR, the sum-throughput of the system is given by ${\Theta _\tau }{\rm{ = }}\sigma \left( {2 - \varepsilon _\tau ^\infty \left( \sigma  \right)} \right)$. By taking the derivative of the expression ${\Theta _\tau }$ with respect to $\sigma$, after some mathematical manipulations and using the identity of the Lambert W function that $Y = X{e^X} \Leftrightarrow X = W\left( Y \right)$ \cite{website}, the optimal information bits that maximize ${\Theta _\tau }$ can be expressed as ${\sigma ^*} = {m \over {\tau \ln 2}}\left[ {W\left( {{{2e} \over A} + e} \right) - 1} \right]$, where $W\left( x \right)$  is the Lambert W function \cite{website} and $A = {{{a_{A,\tau }} + {b_{A,\tau }}} \over {{c_{A,\tau }}}} + {{{a_{B,\tau }} + {b_{B,\tau }}} \over {{c_{B,\tau }}}}$. Furthermore, from (\ref{sumBEPhighSNR}), we can also characterize the minimum value of system blocklength to ensure a specific sum-BLER constraint $\varepsilon$. The minimum blocklength can be solved from (\ref{sumBEPhighSNR}) and given by $\hat m = {{\sigma \tau } \over {{{\log }_2}\left( {1 + {\varepsilon  \over A}} \right)}}$.

In the following, we analytically compare the performance of 2TS scheme and 3TS scheme with optimal values of $\alpha$ and $\beta$ at high SNR range. To this end, we define a sum-BLER ratio of the considered 2TS and 3TS schemes, denoted by ${\varepsilon _{2}^\infty \over \varepsilon _{3}^\infty}$. With the definition, we have 2TS outperforms 3TS in terms of sum-BLER when ${\varepsilon _{2}^\infty \over \varepsilon _{3}^\infty}<1$, and vice versa. Substitute the parameters in Table I, ${\varepsilon _{2}^\infty \over \varepsilon _{3}^\infty}$ can be expressed in (\ref{ratio}) on top of the next page, in which the parameter $\eta$ is given by
\begin{figure*}[!t]
\begin{equation}\label{ratio}
\begin{split}
{\varepsilon _{2}^\infty \over \varepsilon _{3}^\infty} &= {{\left( {{{\bar \gamma }_{A,R}}{\rm{ + }}{{\bar \gamma }_{R,A}}{\rm{ + }}{{\bar \gamma }_{B,R}}} \right){{\bar \gamma }_{A,R}}{{\bar \gamma }_{R,B}} + \left( {{{\bar \gamma }_{B,R}}{\rm{ + }}{{\bar \gamma }_{R,B}}{\rm{ + }}{{\bar \gamma }_{A,R}}} \right){{\bar \gamma }_{B,R}}{{\bar \gamma }_{R,A}}} \over {{{\left( {\sqrt {{{\bar \gamma }_{A,R}}{{\bar \gamma }_{R,A}}\left( {{{\bar \gamma }_{B,R}} + {{\bar \gamma }_{R,B}}} \right)}  + \sqrt {{{\bar \gamma }_{B,R}}{{\bar \gamma }_{R,B}}\left( {{{\bar \gamma }_{A,R}} + {{\bar \gamma }_{R,A}}} \right)} } \right)}^2}}}{{{2^{{{2\sigma } \over m}}} - 1} \over {{2^{{{3\sigma } \over m}}} - 1}}=\left( {1{\rm{ + }}\eta } \right)\left( {{1 \over {1 + {2^{{{2\sigma } \over m}}}/\left( {{2^{{\sigma  \over m}}}{\rm{ + }}1} \right)}}} \right),
\end{split}
\end{equation}
\hrulefill
\vspace*{4pt}
\end{figure*}

\begin{equation}\label{condition2}
\begin{split}
 &\eta = \\
 &{{{\Delta ^2} - {{\bar \gamma }_{R,A}}{{\bar \gamma }_{R,B}}\left( {{{\bar \gamma }_{A,R}} + {{\bar \gamma }_{B,R}}} \right)} \over {{{\left( {\sqrt {{{\bar \gamma }_{A,R}}{{\bar \gamma }_{R,A}}\left( {{{\bar \gamma }_{B,R}} + {{\bar \gamma }_{R,B}}} \right)}  + \sqrt {{{\bar \gamma }_{B,R}}{{\bar \gamma }_{R,B}}\left( {{{\bar \gamma }_{A,R}} + {{\bar \gamma }_{R,A}}} \right)} } \right)}^2}}},
 \end{split}
\end{equation}
$\Delta  = \sqrt {{{\bar \gamma }_{A,R}}{{\bar \gamma }_{R,B}}\left( {{{\bar \gamma }_{A,R}}{\rm{ + }}{{\bar \gamma }_{R,A}}} \right)}  - \sqrt {{{\bar \gamma }_{B,R}}{{\bar \gamma }_{R,A}}\left( {{{\bar \gamma }_{B,R}}{\rm{ + }}{{\bar \gamma }_{R,B}}} \right)} $. It is readily to verify that the term ${{1 \over {1 + {2^{{{2\sigma } \over m}}}/\left( {{2^{{\sigma  \over m}}}{\rm{ + }}1} \right)}}} $ in (\ref{ratio}) is an increasing function of $m$ for $m>0$. We now define a critical value of blocklength $m^*$ such that 2TS and 3TS schemes have the same sum-BLER performance. That is, $m^*$ is a solution to the equation ${\varepsilon _{2}^\infty \over \varepsilon _{3}^\infty} =1$. We can then deduce that for every $m>m^*$, 3TS scheme outperforms 2TS scheme in terms of sum-BLER, and for every $m<m^*$, 2TS scheme outperforms 3TS scheme. In the following, we then aim to solve $m^*$ from ${\varepsilon _{2 }^\infty \over \varepsilon _{3}^\infty} =1$.

With $\sigma >0$ and $m>0$, we can easily prove that $\left( {{1 \over {1 + {2^{{{2\sigma } \over m}}}/\left( {{2^{{\sigma  \over m}}}{\rm{ + }}1} \right)}}} \right)<{2 \over 3}$. Thus, $m^*$ only exists when $\left(1+\eta\right) > { 3 \over 2}$, i.e., $\eta > {1\over2}$. Otherwise, $m^*$ does not exist and ${\varepsilon _{2 }^\infty \over \varepsilon _{3}^\infty}$ is always less than 1. That is, 2TS scheme has a lower sum-BLER than 3TS scheme for any value of $m$ when $\eta \le {1\over 2}$. Based on the above analysis, $m^*$ can be expressed as

\begin{equation}\label{m23}
{m^*} = \left\{ {
\begin{matrix}
\begin{split}
   &{{{\sigma  \over {{{\log }_2}\left( {{{\eta + \sqrt {\eta^2 + 4\eta} } \over {2}}} \right)}}}, \quad \text{if} \quad  \eta > {1\over 2}
   }  \\
   &{\text{Not exist} , \quad \quad \quad \quad\quad \text{if} \quad \eta \le {1\over 2}
   }  \\
\end{split}
 \end{matrix}
 } \right.
\end{equation}

\begin{remark}
The above equation (\ref{m23}) can be used to design a TWRN for short-packet communications under stringent latency requirement. Specifically, let ${m_d}$ be the delay constraint of the network measured in channel uses. We first evaluate the critical value of blocklength $m^*$ based on the expression given in (\ref{m23}) for a specific system setup. If $m^*$ does not exist, we use 2TS scheme for any delay requirement $m_d$. If $m^*$ exists and $m_d>m^*$, we choose 3TS to exchange information between the sources in order to achieve lower sum-BLER. Otherwise when $m_d<m^*$, we should adopt 2TS scheme.

We next examine the impact of different system parameters on the value of $m^*$. From (\ref{m23}), we can see that $m^*$ grows as the transmission information bits $\sigma$ increases. Besides, $m^*$ also increases as $\eta$ decreases or $\Delta$ decreases. The value of $\Delta$ is affected by the average SNR of the four links. For symmetric source transmit powers and relay-to-source distances such that ${{\bar \gamma }_{A,R}} = {{\bar \gamma }_{B,R}}$ and ${{\bar \gamma }_{R,A}} ={{\bar \gamma }_{R,B}}$, we immediately have $\Delta=0$, $\eta < {1\over 2}$ and $m^*$ does not exist. In this scenario, 2TS scheme always outperforms 3TS scheme. For the asymmetric system setup that ${{\bar \gamma }_{A,R}} \ne {{\bar \gamma }_{B,R}}$ or ${{\bar \gamma }_{R,A}} \ne {{\bar \gamma }_{R,B}}$, $\Delta \ne 0$ and $\Delta$ grows as the differences between ${{\bar \gamma }_{A,R}}$ and ${{\bar \gamma }_{B,R}}$,  ${{\bar \gamma }_{R,A}}$ and ${{\bar \gamma }_{R,B}}$ increase. We can now summarize that the critical value of blocklength $m^*$ decreases as the required information bits reduces and the differences between the average SNR of both links increase. We will later verify the above insights by simulation results in the next section.
\end{remark}

%% file: numerical.tex
\section{Numerical Results and Discussions}
In this section, we present some simulation and numerical results to validate and illustrate the above theoretical analysis. In order to capture the effect of path-loss, we use the model ${\Omega _{XY}} = {10^{-3} \over {1 + {d_{XY}^\omega}}}$, where ${\Omega _{XY}}$ with $X,Y \in \left\{A,B,R\right\}$ is the average channel power gain between node $X$ and $Y$, ${d_{XY}}$ denotes the distance between node $X$ and $Y$, and $\omega  \in \left[ {2,5} \right]$ is the path-loss factor. Note that a 30 dB average signal power attenuation is assumed at a reference distance of 1 meter (m) in the above channel model. In all the following simulations, we set the distance between $A$, $B$ and $R$ as ${d_{AB}} = 100$m, $d_{AR}=d_{BR}=50$m, the path-loss factor $\omega  = 3$, and the noise power ${N_A} = {N_B} ={N_R}=- 80$dBm.

\begin{figure}
\centering
  {\scalebox{0.4}{\includegraphics {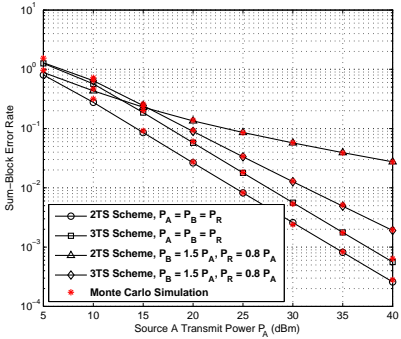}}}
\caption{Sum-BLER versus $P_A$ for different $P_B$ and $P_R$, where $\sigma = 256$ bits and $m=1200$ channel uses.
\label{montsimu}}
\end{figure}

We first compare the analytical expression derived in (\ref{sumBEP}) with the Monte Carlo simulation. The power allocation factors for 3TS scheme are set to be optimal in both the analytical and simulation results. In Fig. \ref{montsimu}, we first can see that the derived analytical expression matches the simulation results well, especially at medium and high SNR range. This verifies the tightness of the approximation we adopted in the derivation. Moreover, for symmetric system setup (i.e., $P_A=P_B=P_R$), 2TS scheme outperforms 3TS scheme which coincides with the analysis provided in Remark 1. For asymmetric system setup, 2TS scheme and 3TS scheme may outperform each other depending on different transmit powers. As the analytical expression agrees well with the simulation, we will only show the analytical results in the following of this section.

\begin{figure}
\centering
  {\scalebox{0.4}{\includegraphics {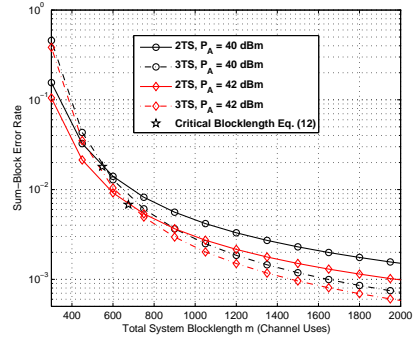}}}
\caption{Sum-BLER versus $m$ for different $P_A$, where $P_B = 55$dBm, $P_R=45$dBm and $\sigma = 1024$ bits.
\label{mplot1}}
\end{figure}


\begin{figure}
\centering
  {\scalebox{0.4}{\includegraphics {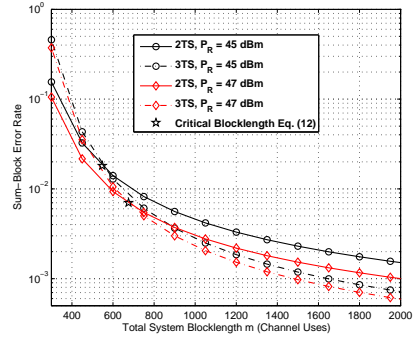}}}
\caption{Sum-BLER versus $m$ for different $P_R$, where $P_A = 40$dBm, $P_B = 55$dBm and $\sigma = 1024$ bits.
\label{mplot3}}
\end{figure}

We then compare the performance of 2TS and 3TS scheme for different total system blocklength $m$ where the power allocation factors in 3TS scheme is set to be optimal. We plot the sum-BLER versus $300 \le m \le 2000$ of 2TS and 3TS schemes for different scenarios. Figs. \ref{mplot1} and Fig. \ref{mplot3} validate our asymptotic analysis and the tightness of the derived critical blocklength in (\ref{m23}). We can observe from Figs. \ref{mplot1}-\ref{mplot3} that when $m<m^*$, 2TS scheme outperforms 3TS scheme and otherwise, 3TS scheme outperforms 2TS scheme. From Fig. \ref{mplot1}, we can observe that as the difference between $P_A$ and $P_B$ grows, the critical value $m^*$ shifts to the left. This means that 3TS scheme is suitable for system with a higher difference between source transmit powers. In Fig. \ref{mplot3}, we see that $m^*$ decreases as $P_R$ reduces, which means that 3TS scheme is appealing to scenario that relay transmit power is relatively low compared with source transmit powers. This is understandable as 3TS allocates different power for different source signals. The less the available power at the relay, the more the significance of this power allocation strategy. All of the above observations coincide well with the analysis provided in Remark 1.